\documentclass[aps,pre,twocolumn,superscriptaddress]{revtex4-1}
\usepackage[dvips]{graphicx}
\usepackage{amsmath}
\begin{document}


\title{Exact solution of a heterogeneous multi-lane asymmetric simple exclusion process}


\author{Takahiro Ezaki}
\affiliation{Department of Aeronautics and Astronautics, Graduate School of Engineering,
The University of Tokyo, 7-3-1 Hongo, Bunkyo-ku, Tokyo 113-8656, Japan}
\author{Katsuhiro Nishinari}
\affiliation{Research Center for Advanced Science and Technology,
The University of Tokyo, 4-6-1, Komaba, Meguro-ku, Tokyo 153-8904, Japan}


\date{\today}

\begin{abstract}
We prove an exact solution of a multi-lane totally asymmetric simple exclusion process (TASEP) with heterogeneous lane-changing rates on a torus. The solution is given by a factorized form; that is, the TASEP in each lane can be separable because of the detailed balance conditions satisfied for lane-changing transitions. Using the saddle point method, the current of particles is calculated in a simple form in a thermodynamic limit. It is notable that the current depends only on a set of lane-changing parameters, not on the configuration of lanes.
\end{abstract}


\maketitle

\section{Introduction}

Driven diffusive systems have been studied actively in recent years, since they are useful for understanding various phenomena in physics and biology.
One of the most important driven particle systems, the totally asymmetric simple exclusion process (TASEP) \cite{ASEP}, was originally proposed as a model for describing biological transport phenomena, 
and has been applied to the modeling of transport processes such as vehicular traffic  \cite{vt}, granular flow  \cite{gf},
and biological transportation by motor proteins  \cite{LK,Kin,Kin2}.

In some studies, the TASEPs with multiple lanes and lane-changing have been investigated analytically \cite{tl,tl2,tl3,tl4,tl6,tl7}, but exact analyses have been performed on a few models \cite{tl7}.

In this work, we consider a multilane system with periodic boundaries in two directions, and present an exact solution in the stationary limit.
The system has $K$ lanes on a cylinder, which is applicable to the problems such as transportation phenomena of the kinesins \cite{Kin,Kin2,Kinl} 
along the 13 protofilaments placed on microtubules cylindrically \cite{Kin3,Kin4}.
On the other hand, when $K=2$ it corresponds to a simple two-lane TASEP with periodic boundary.
Moreover, we do not limit the number of lanes, and thus this work will be a significant achievement for solving a kind of two-dimensional exclusion process exactly.

To construct the solution, we use the detailed balance condition satisfied in lane-changing transitions in the model. For this characteristic, 
the solution has a simple structure, and quantities such as density and current are derived in simple forms.

\section{Model}
We consider a two-dimensional cylindrical lattice of $L\times K$ sites. 
Each lane-$i$ ($i=1,\cdots,K$) has $L$ sites ($j=1,\cdots,L$), and corresponding sites of adjacent lanes are connected with each other.
These $K$ lanes are arranged cylindrically, namely, lane-$i+K$ is identical with lane-$i$.

\begin{figure}[htbp]
 \begin{center}
  \includegraphics[width=50mm]{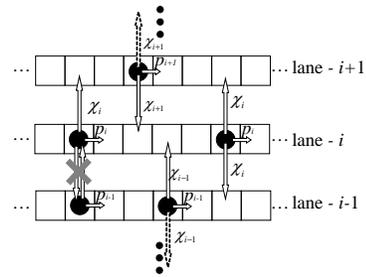}
 \end{center}
 \caption{The schematic view of the system. If the neighboring site to its right is empty,
 the particle hops there $j \rightarrow j+1$ with probability $p_i\Delta t$ (hopping). Moreover, if the receiving site is empty, 
 it changes lanes to the neighboring lane $i+1$ ($i-1$) [$i,j\rightarrow i+1,j$ ($i-1,j$)]
 with probability $\chi_i\Delta t$ (lane-changing). 
 }
 \label{sch}
\end{figure}

A site-($i,j$) can be either be empty ($\tau_{i,j}=0$) or occupied by one particle ($\tau_{i,j}=1$) (the hard-core exclusion). 
$\tau_{i,j}$ denotes the occupation number of the $j$-th site in lane-$i$. The time evolution per time interval $\Delta t$ is written as follows:
\begin{itemize}
\item[(i)]hopping\\$1_{i,j}0_{i,j+1}\rightarrow 0_{i,j}1_{i,j+1}$ with probability $p_i\Delta t$
\item[(ii)]lane-changing\\$1_{i,j},0_{i\pm 1,j}\rightarrow 0_{i,j},1_{i\pm 1,j}$ with probability $\chi_i\Delta t$
\end{itemize}
In each lane, a particle hops to the next site ($j\rightarrow j+1$) with probability $p_i\Delta t(>0)$ if the target site is empty (as in the usual TASEP).
Furthermore, lane-changing transitions are also defined, namely, each particle hops to the adjacent lane ($i\rightarrow i+1$ or $i\rightarrow i-1$) with
probability $\chi_i \Delta t(>0)$ if the target site is empty as shown in FIG.\ref{sch}.
Since in our model each particle is randomly updated, this lane-changing is permitted even if the neighboring site-$j-1$ to the target site is occupied.

When $K=2$ and each lane has only one adjacent lane the lane-changing is restricted to one direction.
This peculiarity is natural and does not influence the argument in the following sections.

\section{Exact solution for a periodic system}
In the following we focus on the periodic system in the hopping direction; namely, site-$j+L$ is identical with site-$j$.
For this system, we present an exact expression for the probability with which a given configuration $\{\tau_{i,j}\}$ is realized, as a product form of 
the density weight $\Psi_i(M_i)$ and the configuration weight in lane-$i$, $g_i(\{\tau_{i,j}\}_i)$. Here, $M_i$ is the number of particles in lane-$i$ ($M_i=\sum_{j}{\tau_{i,j}}$) and $\{\tau_{i,j}\}_i=\tau_{i,1},\tau_{i,2},\cdots \tau_{i,L}\in \{0,1\}$ is a set of occupation 
numbers in lane-$i$.
\begin{eqnarray}
P(\{\tau_{i,j}\})&=&Z_{L,N,K}^{-1} \prod_{i=1}^{K}{\Psi_i(M_i) g_i(\left\{ \tau_{i,j}\right\}_i)}\label{so1}\\
&=&Z_{L,N,K}^{-1} \prod_{i=1}^{K}{\left(\frac{1}{\chi_i} \right)^{M_i}}\label{so2}
\end{eqnarray}
Here, each weight factor $g_i(\{\tau_{i,j}\}_i)$ is assumed to be $1$ as in the usual ASEP with periodic boundary \cite{sg},
and the density weight $\Psi_i(M_i)$ is defined as $\Psi_i(M_i)=(1/\chi_i)^{M_i}$.
The normalization factor $Z_{L,N,K}$ is thus written as
\begin{equation}
Z_{L,N,K}=\sum_{M_1=0}^L {\scriptstyle\cdots} \sum_{M_K=0}^L \prod_{i=1}^{K} \left(\frac{1}{\chi_i}\right)^{M_i}\left(
\begin{array}{c}
L  \\
M_i 
\end{array}
\right)
\delta \left(\sum_{i=1}^{K}{M_i} - N \right)\label{Z}
\end{equation}
by taking the sum of the weights with respect to $M_i$. The $\delta$-function ensures that only those configurations with the correct total number of particles are included.

Next, we confirm that this expression gives the exact solution.
The exact solution for the steady state of the system must satisfy the master equation described below, and conversely, the expression satisfying the master equation must be the exact solution.

\begin{equation}
0=\frac{\partial }{\partial t}P(\mathcal{C}) = \sum_{\mathcal{C'}\neq \mathcal{C}}{\left\{P(\mathcal{C'})W(\mathcal{C'}\rightarrow \mathcal{C})-P(\mathcal{C})W(\mathcal{C}\rightarrow \mathcal{C'})\right\}}\label{mas}
\end{equation}
Here, $\mathcal{C}$ and $W$($\mathcal{C}\rightarrow\mathcal{C'}$) indicates the configuration of particles and the transition probability from configuration $\mathcal{C}$ to $\mathcal{C'}$ respectively. 

We separate the transitions into two parts according to their type of motion, i.e., hopping or lane-changing. It is obvious that the terms for the hopping transition in the master equation (\ref{mas}) vanish 
when one substitutes the presented solution, since each $g_i(\{\tau_{i,j}\}_i)$ is the exact solution of the TASEP with periodic boundary in each lane. For transition terms of hopping in lane-$i$, one only has to consider
the weight of lane-$i$ $g_i(\{\tau_{i,j}\}_i)$ since other terms in Eq. (\ref{so1}) are in common before and after the transition.

\begin{figure}[htbp]
 \begin{center}
  \includegraphics[width=60mm]{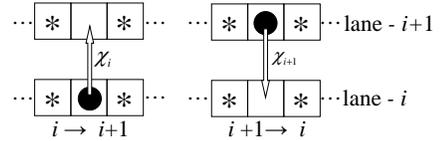}
 \end{center}
 \caption{Correspondence of lane-changing transitions on a certain site in lanes $i$ and $i+1$. Asterisks indicate the same configuration in both transitions.}
 \label{cor}
\end{figure}

\begin{figure*}[htbp]
 \begin{center}
  \includegraphics[width=140mm]{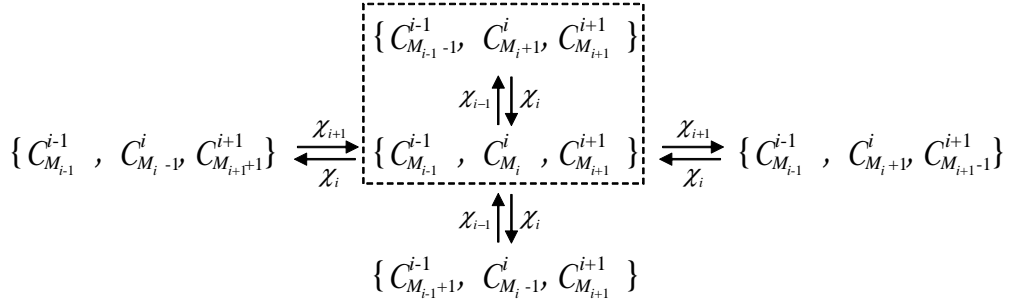}
 \end{center}
 \caption{Possible lane-changing transitions concerning lane-$i$ for an arbitrary set of configurations $\{\mathcal{C}^{i-1}_{M_{i-1}},\mathcal{C}^i_{M_{i}},\mathcal{C}^{i+1}_{M_{i+1}}\}$. $\mathcal{C}^i_{M_{i}}$ represents a certain configuration in lane-$i$, where $M_i$ is the total number of particles in lane-$i$ in the configuration.
 }
 \label{trans}
\end{figure*}

Then we show that the rest of Eq. (\ref{mas}), namely, 
the terms for lane-changing transitions, also vanish. We focus on the lane-changing transitions concerned with lane-$i$. It is sufficient to consider the neighboring lanes ($i-1,i,i+1$), and we write the configurations of
these three lanes with $M_{i-1}, M_{i}$ and $M_{i+1}$ particles as $\{\mathcal{C}^{i-1}_{M_{i-1}},\mathcal{C}^i_{M_{i}},\mathcal{C}^{i+1}_{M_{i+1}}\}$. First, let us confirm the correspondence of 
the lane-changing transitions $i\rightarrow i+1$ and $i+1\rightarrow i$. As shown in FIG.\ref{cor}, the number of configurations of these transitions are the same where the asterisks indicate the common configurations 
among the lanes, namely, every lane-changing transition $\{\cdots ,\tau_{i,j}=1,\cdots,\tau_{i\pm 1,j}=0,\cdots\}\rightarrow\{\cdots ,\tau_{i,j}=0,\cdots,\tau_{i\pm 1,j}=1,\cdots\}$ is paired with its counterpart 
$\{\cdots,\tau_{i,j}=0,\cdots,\tau_{i\pm 1,j}=1,\cdots\}\rightarrow\{\cdots ,\tau_{i,j}=1,\cdots,\tau_{i\pm 1,j}=0,\cdots\}$.
These corresponding transitions balance in the master equation as explained below. Here, we avoid the discussion with explicit expressions expanding (\ref{mas})
 because the expressions  would be unnecessarily complicated and make us lose sight of the essence.
The balance of lane-changing transitions concerned with lane-$i$ is illustrated in FIG.\ref{trans}.
The transitions not described in the figure are forbidden in this model (they occur with probability $0$). 
We choose one arbitrary configuration of lanes $i-1, i$, and $i+1$, $\{\mathcal{C}^{i-1}_{M_{i-1}},\mathcal{C}^i_{M_{i}},\mathcal{C}^{i+1}_{M_{i+1}}\}$. Taking the transition between 
 $\{\mathcal{C}^{i-1}_{M_{i-1}},\mathcal{C}^i_{M_{i}},\mathcal{C}^{i+1}_{M_{i+1}}\}$ and  $\{\mathcal{C}^{i-1}_{M_{i-1}-1},\mathcal{C}^i_{M_{i}+1},\mathcal{C}^{i+1}_{M_{i+1}}\}$ (in the broken line in the figure)
 as an example, they cancel in the master equation as follows:
\begin{eqnarray}
&&P(\mathcal{C'})W(\mathcal{C'}\rightarrow \mathcal{C})-P(\mathcal{C})W(\mathcal{C}\rightarrow \mathcal{C'})\\
&=&Z^{-1}_{L,N,K}\prod_{\substack{1\leq l\leq L\\l\neq i-1,i,i+1}}\left(\frac{1}{\chi_l}\right)^{M_l}\nonumber\\
                  &\times&\biggl\{ \biggr. \left(\frac{1}{\chi_{i-1}}\right)^{M_{i-1}-1}\left(\frac{1}{\chi_{i}}\right)^{M_{i}+1}\left(\frac{1}{\chi_{i+1}}\right)^{M_{i+1}} \chi_i\nonumber\\
                  &&\quad-\left(\frac{1}{\chi_{i-1}}\right)^{M_{i-1}}\left(\frac{1}{\chi_{i}}\right)^{M_{i}}\left(\frac{1}{\chi_{i+1}}\right)^{M_{i+1}}\chi_{i-1}\biggl. \biggr\}\\
                  &=&0
\end{eqnarray}
One can understand the rest of transitions also balance using the same argument. 
Since the system is periodic, this balance holds for every lane and configuration. It should be noted that the transitions between the neighboring lanes satisfy the detailed balance condition. 
Thus, we have proved that the presented expression surely satisfies the master equation and correctly describes the system.

\section{Thermodynamic Limit}
Let us discuss the thermodynamic limit ($L\rightarrow \infty$) of the system. First, we consider the following functions.
\begin{eqnarray}
f_i(M) &=& \left(\frac{1}{\chi_i}\right)^M \left(
\begin{array}{c}
L  \\
M 
\end{array}
\right),\\
\mathcal{Z}_{L,K}(s) &=& \sum_{M_1=0}^L \cdots \sum_{M_K=0}^L \prod_{i=1}^{K}f_i(M_i)s^{M_i}\\
&=& \prod_{i=1}^K \sum_{M_i=0}^{L}f_i(M_i)s^{M_i}\\
&=& \prod_{i=1}^K \left( 1+\frac{s}{\chi_i}\right)^L\\
&=& \left[ \prod_{i=1}^K \left( 1+\frac{s}{\chi_i}\right)  \right]^L = \left[F(s)\right]^L
\end{eqnarray}
where $f_i(M)$ is the weight of lane-$i$ and $F(s)=\prod^K_{i=1}{(1+s/\chi_i)}$. Using this $\mathcal{Z}_{L,K}(s)$, the partition function $Z_{L,N,K}$ is expressed
in an integral form.
\begin{equation}
Z_{L,N,K} = \oint \frac{ds}{2\pi i}\frac{\mathcal{Z}_{L,K}(s)}{s^{N+1}} =\oint \frac{ds}{2\pi i}\frac{\left[F(s)\right]^L}{s^{N+1}}\label{ZZ}
\end{equation}
We evaluate the integral (\ref{ZZ}) in the $L\rightarrow \infty$ limit, keeping $N/L=\rho$($<K$)  constant, by the saddle point method.
For large $L,N$ Eq. (\ref{ZZ}) is dominated by the saddle point of the integral denoted by $s=z$.
Following  \cite{sg}, we define 
\begin{equation}
\phi(s) = -\rho \ln s + \ln [F(s)].
\end{equation}
Then the saddle point is given by $\phi'(z)=0$ as,
\begin{equation}
\rho = \sum_{i=1}^K{\frac{z/\chi_i}{1+z/\chi_i}}\label{r}.
\end{equation}
Moreover, the partition function is evaluated by considering the thermodynamic limit of Eq. (\ref{ZZ}) using this saddle point as,
\begin{eqnarray}
Z_{L,N,K}&\simeq& \frac{1}{(2\pi L)^{1/2}}\frac{1}{\vert\phi''(z)\vert^{1/2}}\frac{\exp{\left(L\phi(z)\right)}}{z}\nonumber\\
&=& \frac{1}{(2\pi L)^{1/2}}\frac{1}{\vert\phi''(z)\vert^{1/2}}\frac{\mathcal{Z}_{L,K}(z)}{z^{N+1}}
\end{eqnarray}
Let us investigate the current in lane-$i$ which is defined as $J_i=\left\langle \tau_{i,j}(1-\tau_{i,j+1})p_i\right\rangle$, by considering 
the configurations where $\tau_{i,j}=1,\tau_{i,j+1}=0$. The weight of these configurations is calculated through its generating function,
\begin{eqnarray}
\tilde{\mathcal{Z}}_{L,K}^{i}(s) &=& \prod_{i'\neq i}^K\left(\sum_{M_{i'}=0}^{L}f_{i'}(M_{i'})s^{M_{i'}}\right)\nonumber\\
&&\times \sum_{M_i=0}^L\left(\frac{1}{\chi_i}\right)^{M_i} \left(
\begin{array}{c}
L-2  \\
M_i-1 
\end{array}
\right)s^{M_i}\nonumber\\
&=& \frac{s/\chi_{i}}{(1+s/\chi_{i})^2}\left[F(s)\right]^L.
\end{eqnarray}
Therefore, by the same argument we find
\begin{eqnarray}
J_i &=& \frac{\tilde{Z}^i_{L,N,K}}{Z_{L,N,K}}p_i\\
&=&\frac{\oint \frac{ds}{2\pi i}\frac{s/\chi_i}{(1+s/\chi_i)^2}\frac{\left[F(s)\right]^L}{s^{N+1}}}{\oint \frac{ds}{2\pi i}\frac{\left[F(s)\right]^L}{s^{N+1}}}\\
&\simeq& \frac{z/\chi_i}{(1+z/\chi_i)^2}p_i\label{cur}
\end{eqnarray}  
in the thermodynamic limit. Here, $z$ is the saddle point of the integral again. Note that the two integrals performed above
have common $\phi$, and thus the saddle point is consistent. 
In a similar way, the density in lane-$i$ $\rho_i=M_i/L$ is also calculated as,
\begin{equation}
\rho_i \simeq \frac{z/\chi_i}{1+z/\chi_i}.
\end{equation} 
This corresponds with the expression (\ref{r}) where each $\rho_i$ contributes to the total density.
Furthermore, it is quite notable that $z$ plays the role of a ``common incoming rate" when we compare it with the Langmuir equilibrium density $\kappa/(1+\kappa)$ where $\kappa$ is the
ratio of the attachment and detachment rates \cite{LK}.

\begin{figure}[htbp]
 \begin{center}
  \includegraphics[width=80mm]{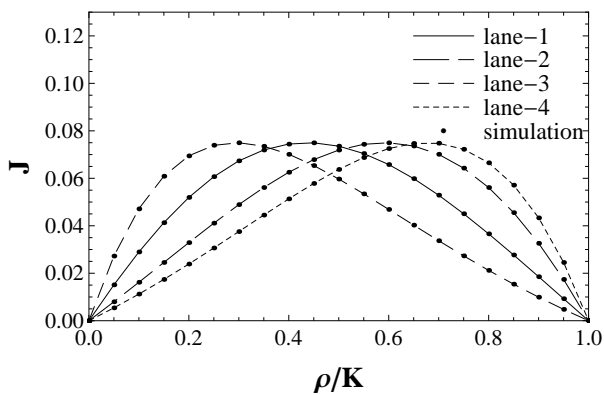}
 \end{center}
 \caption{Fundamental diagram for the system with four lanes, $L=1000$, $(\chi_1,\chi_2,\chi_3,\chi_4)=(0.1,0.2,0.05,0.3)$ and $p_1=p_2=p_3=p_4=0.3$. Here, the hopping probabilities
 $p_i$ are set as constant to make the peaks of each curve aligned for the comprehensibility.  
 The relationship of the curves depends on the lane-changing probability, and the relatively large $\chi_i$ leads to the large critical density. 
 }
 \label{graph}
\end{figure}

The currents $J_i$ obtained from Monte Carlo simulations are plotted on FIG. \ref{graph} with theoretical lines. 
The theoretical lines are obtained from Eq. (\ref{cur}) after one finds $z$ from Eq. (\ref{r}).
It is notable that the density in lane $i$ depends only on the set of lane-changing rates, and is independent of the configuration of the lanes.
For relatively large leaving rate of lane-$i$ $\chi_i$ the density becomes small, and leads to the large  critical density.

To summarize this discussion, we can also regard the dynamics as the ASEP with Langmuir kinetics \cite{LK} on each lane with detachment rate $\chi_i$ and
 effective attachment rate $z$ in the thermodynamic limit for its detailed balance property.

\section{Summary}
In this work we have considered a cylindrical multi-lane exclusion process and presented an exact solution of it in the stationary limit.
This solution is also applicable to the model with $K=2$ lanes, which corresponds to a two-lane model often discussed in transportation problems.
Using the saddle point method we have derived an expression for the current and density, and have shown that simulation results with representative parameters well agree with the theory.
As shown in FIG.\ref{graph}, the peak shift has been observed, and this phenomenon might be seen in the actual transport process such as the biological transportation of motor proteins.

The most important feature of the model is the detailed balance condition satisfied in the lane-changing transitions; and for this characteristic, 
the solution has been constructed simply and interpreted as separated ASEPs with Langmuir kinetics with a common rate in the thermodynamic limit.

It is also intriguing that the system is solvable although the one-dimensional heterogeneous symmetric simple exclusion process (which corresponds to the 
dynamics in the $i$ direction in this work)
itself has not been solved in the previous works so far. Moreover, it is significant that an exact solution for a two-dimensional exclusion process has been given.

In the model we have considered a system with symmetric lane-changing rates, where decrease of particles in the lane corresponds to one lane-changing parameter.
If we assume asymmetric ones, the formulation of the expression would be more complex, and it should be investigated in future works.

\bibliography{ref}

\begin{thebibliography}{00}
\bibitem{ASEP}J. T. MacDonald, J. H. Gibbs, and A. C. Pipkin, Biopolymers 6, 1 (1968).
\bibitem{vt}D. Chowdhury, L. Santen, A. Schadschneider, Phys. Rep. 329 (2000) 199.
\bibitem{gf}H. Hayakawa, K. Nakanishi, Prog. Theor. Phys. Suppl. 130 (1998) 57.
\bibitem{LK}A. Parmegianni, T. Franosh, E. Frey, Phys. Rev. Lett. 90, 086601 (2003).
\bibitem{Kin}K. Nishinari, Y. Okada, A. Schadschneider, and D. Chowdhury, Phys. Rev. Lett. 95, 118101 (2005).
\bibitem{Kin2}P. Greulich, A. Garai, K. Nishinari, A. Schadschneider, and D. Chowdhury, Phys. Rev. E 75, 041905 (2007).
\bibitem{tl}V. Popkov and I. Peschel, Phys. Rev. E 64, 026126 (2001).
\bibitem{tl2}H.-W. Lee, V. Popkov, and D. Kim, J. Phys. A 30, 8497 (1997).
\bibitem{tl3}R. Jiang, M.-B. Hu, Y.-H. Wu, and Q.-S. Wu, Phys. Rev. E 77, 041128 (2008).
\bibitem{tl4}E. Pronina and A. B. Kolomeisky, J. Phys. A 37, 9907 (2004).
\bibitem{tl6}T. Mitsudo and H. Hayakawa, J. Phys. A 38, 3087 (2005).
\bibitem{tl7}M. Kanai, Phys. Rev. E 82, 066107 (2010).
\bibitem{Kinl}D. Chowdhury, A. Garai, and J. S. Wang, Phys. Rev. E 77, 050902(R) (2008).
\bibitem{Kin3}J. Howard, \textit{Mechanics of Motor Proteins and the Cytoskeleton} (Sinauer Associates, Sunderland, 2001).
\bibitem{Kin4}\textit{Molecular Motors}, edited by M. Schliwa (Wiley-VCH, New York, 2002).
\bibitem{sg}R. A. Blythe and M. R. Evans, J. Phys. A 40, R333 (2007).

\end{thebibliography}

\end{document}